# Hybrid Physical and Geometrical Optics Method for Modeling Subsurface Imaging Using mmWave FMCW Radar

Kaito Ichijo, Hang Song, Xin Du, Bo Wei, and Junichi Takada

*Abstract*— A hybrid physical and geometrical optics method is proposed to model the subsurface imaging using mmWave FMCW radar. Modeling of the wave propagation for subsurface imaging can improve the interpretation of acquired data and imaging results. Full-wave simulation is common in simulating wave propagation. However, when the frequency is high such as mmWave frequency, it is difficult to implement since it costs large computation resource and time. In this paper, the physical and geometrical optics are hybridized to simulate the wave propagation in subsurface imaging scenarios. In the proposed method, physical optics method is utilized to calculate the reflection from the object and geometrical optics method is utilized to calculate the transmission of the wave through object. By combining the results from physical and geometrical optics, the wave propagation in the subsurface imaging scenarios is simulated. The synthetic-aperture radar imaging is applied to the simulated data and the image is successfully reconstructed. Further, the experiment setup is developed and the comparison between simulation and experiment is carried out. The results demonstrated that the proposed simulation method can model the subsurface imaging with mmWave FMCW radar.

## I. INTRODUCTION

SUBSURFACE imaging is utilized for a wide range of applications such as through-the-wall surveillance [1], vital signal monitoring [2], non-destructive infrastructure inspection [3], and security check [4]. When determining the health status of infrastructure due to deterioration with aging, abnormalities that can be seen from the outside such as exterior walls or power lines can be detected via visual methods [5]. However, when inspecting the interior structures or defects which are covered by other material such as crack detection [6] and pipeline inspection [7], non-destructive inspection methods are necessary to avoid damaging the original objects.

There exist many works which developed subsurface imaging techniques for different purposes. González-Partida *et al.* theoretically analyzed the feasibility of using millimeter wave (mmWave) frequency modulated continuous wave (FMCW) radar for through-the-wall surveillance [1]. The wall attenuation and reflections were estimated with analytical models. The experiment with 34 GHz radar was conducted to demonstrate its validity. Ghasr *et al.* presented a high-resolution imaging system with a linear one-dimensional imaging array operating at 30 GHz for non-destructive testing [8]. With the proposed system, a rubber inserted in wood was detected. The flaws on radome were also detected via imaging. Gao *et al.* utilized the mmWave reflectometry and imaging for diagnosing skin burn injuries non-invasively [9]. The performance was evaluated at V-band (50-75 GHz). Becquaert et al. proposed a novel method with compressed sensing for stepped frequency continuous wave radar through-the-wall imaging [10]. The performance was validated at the frequency range 1 ~ 5 GHz. Yektakhah *et al*. presented an all-directions through-the-wall imaging system by using a dense 2-D synthetic array and bi-static FMCW transceivers [11]. The performance was evaluated at the frequency 1.2 ~ 1.968 GHz which provided images of walls and objects. Pramudita *et al*., presented a radar system to detect the vital sign of human behind wall with 24 GHz FMCW radar [12]. The method to exclude the beat signal from wall obstacle was proposed and the experiment result showed the efficacy. Naghavi *et al*. presented a 100-GHz FMCW radar for imaging and the performance was verified by inspecting the drywall [13]. Tian *et al*. developed the traveling wave tubes for enhancing the penetration ability of through-the-wall radar system, operating at 93.6 ~ 96 GHz [14]. Yanik *et al*. developed a mmWave imaging testbed using 79 GHz FMCW radar [15].

Although various imaging systems have been developed at different working frequencies, mmWave FMCW system shows the advantages that it can be integrated with more compact size and provides higher resolution owing to shorter wavelength. On the other hand, the attenuation becomes larger with the increase of frequency. To predict the performance, simulation is essential to evaluate the proposed system before application and compare with the experiment results. Modeling of the wave propagation for subsurface imaging can improve the interpretation of acquired signals and imaging results.

Many simulation methods are utilized to simulate wave propagation. Hirata *et al*. proposed a non-destructive system for detecting concrete surface cracks covered by paper [16].

Kaito Ichijo and Junichi Takada are with the Department of Transdisciplinary Science and Engineering, Institute of Science Tokyo, Tokyo 152-8550, Japan.

Hang Song is with the Research Institute for Semiconductor Engineering, Hiroshima University, Higashi-Hiroshima 739-8527, Japan. He was with the Department of Transdisciplinary Science and Engineering, Institute of Science Tokyo, Tokyo 152-8550, Japan. (e-mail: hanghsong@hiroshima-u.ac.jp)

Xin Du is with Graduate School of Science and Engineering, Kagoshima University, Kagoshima-shi, Kagoshima 890-8580, Japan.

Bo Wei is with School of Informatics and Data Science, Hiroshima University, Higashi-Hiroshima 739-8527, Japan, and also with Japan Science and Technology Agency (JST), PRESTO, Kawaguchi, Saitama 332-0012, Japan.

The mmWave imaging unit operating at 76.5 GHz was utilized to scan the concrete and the image was created to recognize the crack. Simulation based on finite element method was carried out to analyze the dependance of image contrast on paper thickness. Liu *et al.* utilized the finite difference time domain (FDTD) to investigate the radar system for through-the-wall vital signal extraction and bio-radiolocation [17]. The modeled radar system operates at 1 GHz. Gonzalez-Valdes *et al.* presented the mmWave multistatic radar imaging system for human body screening [4]. The simulations both in 2D and 3D were utilized to validate the system at 23 ~ 28 GHz. In 2D simulation, curved metallic surface resembling the torso of the human body was modeled and the method of moments was utilized. In 3D simulation, the physical optics (PO) method was utilized. Helander *et al.* presented mmWave nondestructive testing of the defects on composite panels using compressive sensing [18]. The frequency range was 50 ~ 67 GHz. Simulations were conducted for proof of concept of the proposed algorithms. Although the simulation method was not specified, FEKO was used. Zhang *et al.* developed a mmWave holographic imaging radar system for nondestructive testing. The radar operates at 195 GHz and the bandwidth is 45 GHz [19]. The FDTD method was used to validate the imaging system. Aljurbua *et al.* presented a 3D bistatic subsurface imaging radar for detecting and localizing pipeline leaks [20]. The frequency-domain full-wave simulation with FEKO was carried out in the range 200 ~ 500 MHz with a step of 6 MHz. Eide *et al.* modeled the FMCW radar using FDTD method for analyzing the subsurface [21]. The center frequency of 675 MHz and bandwidth of 1050 MHz were considered.

Although full-wave methods both in time and frequency domains are common in simulating the wave propagation, it costs large computation resource and time for modeling the mmWave FMCW imaging system since the frequency is high, the signal duration is long in relatively large problem size, and scanning is necessary. High-frequency asymptotic approximations such as PO can reduce the computation time, but the influence of the obstacle such as wall is not included, and the target is generally modeled as metallic material.

In this research, a hybrid physical and geometrical optics (HPGO) method is proposed for modeling subsurface imaging with mmWave FMCW radar. The proposed method aims at providing a more suitable simulation approach for investigating the subsurface imaging in high frequency range. It takes advantage of the PO for simulating the reflections and back scattering from the objects. Since both the obstacle and target reflect the waves, the PO is applied to simulate the scattering from both obstacle and target. To simulate the refraction at the boundary of the obstacle and the transmission features through the blockage, geometrical optics (GO) is utilized. By combining the results from PO and GO, the wave propagation in the scenarios such as the through-the-wall is approximated. Compared with the full wave simulation, the proposed method can reduce the computation resource. Meanwhile, it can simulate the influence of the obstacles which can be applied to subsurface imaging. To evaluate the performance of the proposed method, the experiment system is also developed. By comparing the results from simulation and experiment, it demonstrates that the proposed HPGO can model the characteristics of mmWave FMCW radar for subsurface imaging.

The remainder of this paper is organized as follows. Section II presents the details of the proposed HPGO method. Section III explains the system development by using the FMCW radar for imaging and the corresponding simulation configuration. Section IV presents the simulation results and the comparison with measurement. Finally, the conclusion is made in Section V.

## II. HYBRID PHYSICAL AND GEOMETRICAL OPTICS METHOD

The model shown in Fig. 1 is considered for simulating the subsurface imaging problem with mmWave FMCW radar. The model consists of a monostatic radar, a dielectric slab which mimics any obstacle such as wall, a target which mimics the object to be detected. To generate the image, scanning is utilized in the operation, and the radar moves along the $x$ axis. In the proposed HPGO method, the reflection from the obstacle is calculated by PO method [22]. When the wave propagates through the obstacle, the influence of the obstacle is estimated by GO. Sequentially, the reflection from the target is calculated by PO, and the influence from the obstacle is also included. In this method, the obstacle and target can be either metallic or dielectric materials.

In order to synthesize the data for imaging, the simulation is carried out at each scanning point. The coordinate of the radar

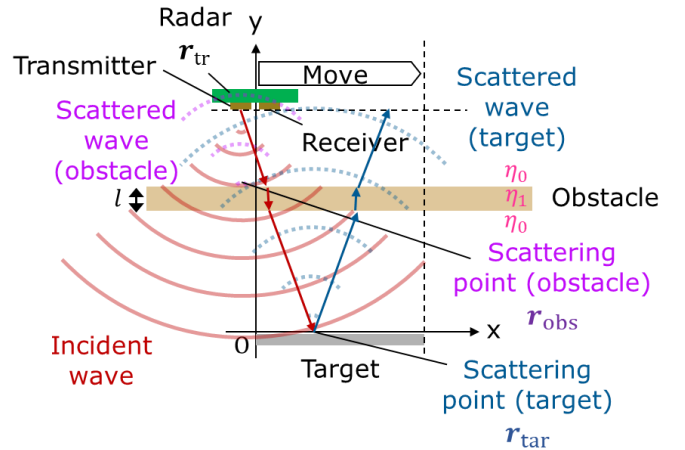

Fig. 1. Schematic diagram of HPGO for simulating wave propagation in subsurface imaging.

at $x$ axis is defined as $x_{\text{tr}}$. The obstacle and target are placed along the $x$ axis. The positions of the target, obstacle, and the radar at $y$ axis are defined as 0, $y_{\text{obs}}$, and $y_{\text{tr}}$, respectively. For any scanning point $\boldsymbol{r}_{\text{tr}}(x_{\text{tr}}, y_{\text{tr}})$, the scattering from the obstacle is expressed as follows:

$$\boldsymbol{E}_{\text{obs}}(\boldsymbol{r}_{\text{tr}}) = \int [-j\omega\mu_0 \cdot \boldsymbol{J}_{\text{obs}}(\boldsymbol{r}_{\text{obs}}) \cdot G(\boldsymbol{r}_{\text{tr}}, \boldsymbol{r}_{\text{obs}}) + \boldsymbol{M}_{\text{obs}}(\boldsymbol{r}_{\text{obs}}) \times G(\boldsymbol{r}_{\text{tr}}, \boldsymbol{r}_{\text{obs}})]\mathrm{d}\boldsymbol{r}_{\text{obs}} \qquad (1)$$

where $\boldsymbol{r}_{\text{obs}}$ is any point on the obstacle. $\omega$ is the angular frequency of the incident wave, and $\mu_0$ is the magnetic permeability of air. $G(r_{tr}, r_{obs})$ is the Green's function. $J_{obs}(r_{obs})$ and $M_{obs}(r_{obs})$ are the equivalent electric current and equivalent magnetic current at $r_{obs}$ on the obstacle, respectively. They are calculated as:

$$J_{obs}(r_{obs}) = n \times H(r_{obs}) \quad (2)$$

$$M_{obs}(r_{obs}) = E(r_{obs}) \times n \quad (3)$$

where $E(r_{obs})$ and $H(r_{obs})$ are the electric field and magnetic field at $r_{obs}$ on the obstacle, respectively. $n$ is the normal vector of the obstacle. Since PO approximates the total field as the superposition of the incident wave and reflected wave, (2) and (3) can be further expressed as:

$$J_{obs}(r_{obs}) = (1 - R_{obs})n \times H_i(r_{obs}) \quad (4)$$

$$M_{obs}(r_{obs}) = (1 + R_{obs})E_i(r_{obs}) \times n \quad (5)$$

where $E_i(r_{obs})$ and $H_i(r_{obs})$ are the incident fields propagating at $r_{obs}$. $R_{obs}$ is the reflection coefficient of the obstacle which is calculated as:

$$R_{obs} = \frac{\cos\theta_1^i(r_{obs}) - \sqrt{\frac{\varepsilon_1}{\varepsilon_0}}\sqrt{1 - \left(\frac{\varepsilon_0}{\varepsilon_1}\right)\sin^2\theta_1^i(r_{obs})}}{\cos\theta_1^i(r_{obs}) + \sqrt{\frac{\varepsilon_1}{\varepsilon_0}}\sqrt{1 - \left(\frac{\varepsilon_0}{\varepsilon_1}\right)\sin^2\theta_1^i(r_{obs})}} \quad (6)$$

where $\varepsilon_0$ and $\varepsilon_1$ are the complex permittivities of the air and obstacle, respectively. $\theta_1^i(r_{obs})$ is the incident angle of the wave at $r_{obs}$ of the obstacle. Here, the material of the obstacle and target are considered non-magnetic, and hence their permeabilities are the same.

The cylindrical wave is considered, and the two-dimensional Green's function is utilized as:

$$G(r_{tr}, r_{obs}) = -\frac{1}{4j}H_0^{(2)}(k|r_{tr} - r_{obs}|) \quad (7)$$

where $k$ is the wave number. $H_0^{(2)}$ is the second kind of Hankel function for the zeroth order. Electromagnetic waves attenuate in amplitude with distance $|r_{tr} - r_{obs}|$. This attenuation is expressed using the Green's function. The electric field at the transmitter is defined as $E_0$. The electric field of the incident wave at $r_{obs}$ can be written as:

$$E_i(r_{obs}) = E_0 H_0^{(2)}(k|r_{tr} - r_{obs}|) \quad (8)$$

The magnetic field of the incident wave at $r_{obs}$ can be written as:

$$H_i(r_{obs}) = \hat{s} \times \frac{E_i(r_{obs})}{\eta_0} \quad (9)$$

where $\hat{s}$ is the normal vector of the direction of wave propagation. $\hat{s}$ is calculated as:

$$\hat{s} = \frac{(r_{obs} - r_{tr})}{|r_{tr} - r_{obs}|} \quad (10)$$

$\eta_0$ is the impedance of air, and it is calculated as $\sqrt{\mu_0/\varepsilon_0}$.

When estimating the reflection from target which is shadowed by the obstacle, the influence of the obstacle should be included. The electromagnetic wave changes when penetrating the obstacle between the radar and the target object. The obstacle is considered sufficiently large, such as walls of the building, compared with the wavelength of the FMCW radar. The surface of the obstacle is assumed to be smooth in this research. Under these conditions, the electromagnetic wave can be approximated as that it travels straight in air and refracts on the boundary between two different materials with different dielectric properties. Here, the estimation of reflection from target is also based on the PO method. But compared the case of the obstacle, additional terms are implemented to the amplitude and phase of the electric field for compensating the effect of obstacle. The additional terms are calculated by geometrical optics (GO) as shown in Fig. 2.

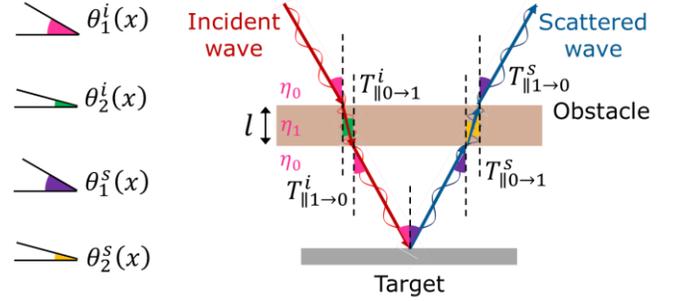

Fig. 2. Schematic diagram of GO for simulating wave transmission inside the obstacle.

When wave propagates at the boundary between the air and obstacle, transmission occurs from air to the obstacle and the transmission coefficient is defined as $T_{\|0 \to 1}^i$. Then the wave propagates inside the obstacle. When the wave propagates at the boundary between the obstacle and air, transmission occurs from the obstacle to air and the transmission coefficient is defined as $T_{\|1 \to 0}^i$. These two coefficients are calculated as:

$$T_{\|0 \to 1}^i = \frac{2\eta_1 \cos\theta_1^i}{\eta_0 \cos\theta_1^i + \eta_1 \cos\theta_2^i} \quad (11)$$

$$T_{\|1 \to 0}^i = \frac{2\eta_0 \cos\theta_2^i}{\eta_1 \cos\theta_2^i + \eta_0 \cos\theta_1^i} \quad (12)$$

where $\eta_1$ is the impedance of air. $\eta_1$ is calculated as $\sqrt{\mu_1/\varepsilon_1}$. $\theta_1^i$ and $\theta_2^i$ are the incident angle and the refraction angle of the wave when propagating from air to the obstacle, respectively. The relation between $\theta_1^i$ and $\theta_2^i$ is depicted as:

$$\theta_2^i = \sin^{-1}\left(\sin\theta_1^i \sqrt{\frac{\varepsilon_0}{\varepsilon_1}}\right) \quad (13)$$

After the wave propagates through the obstacle, the attenuation is added by multiplying $T_{\|0 \to 1}^i$ and $T_{\|1 \to 0}^i$. Since the wavelength changes when the wave enters the dielectric obstacle, the additional phase change is included which can be calculated as:

$$\varphi_i = k_{obs}l\left(\frac{1}{\cos\theta_1^i} - \frac{1}{\cos\theta_2^i}\sqrt{\frac{\varepsilon_1}{\varepsilon_0}}\right) \quad (14)$$

where $k_{obs}$ is the wave number in the obstacle. $l$ is the thickness of the obstacle.

When the wave propagates at the target and scatters, the scattered wave will propagate back to the receiver. Similar to the incident wave, when the wave propagates at the boundary between the air and obstacle, transmission occurs from the air to obstacle, and the transmission coefficient is defined as $T_{\parallel 0 \to 1}^{S}$. Then the transmission coefficient when the wave propagates from obstacle to air is defined as $T_{\parallel 1 \to 0}^{S}$. These two coefficients are calculated as:

$$T_{\parallel 0 \to 1}^{S} = \frac{2\eta_1 \cos \theta_1^S}{\eta_0 \cos \theta_1^S + \eta_1 \cos \theta_2^S} \quad (15)$$

$$T_{\parallel 1 \to 0}^{S} = \frac{2\eta_0 \cos \theta_2^S}{\eta_1 \cos \theta_2^S + \eta_0 \cos \theta_1^S} \quad (16)$$

where $\theta_1^S$ and $\theta_2^S$ are the incident and refraction angles of the scattered wave when the wave propagates from air to the obstacle, respectively. The additional phase change caused by the obstacle for the scattered wave is calculated as:

$$\varphi_s = k_{\text{obs}} l \left( \frac{1}{\cos \theta_1^S} - \frac{1}{\cos \theta_2^S} \sqrt{\frac{\varepsilon_1}{\varepsilon_0}} \right) \quad (17)$$

Taking consideration of attenuation and phase change during the transmission inside the obstacle, the reflection from the target is approximated as follows:

$$\boldsymbol{E}_{\text{tar}}(\boldsymbol{r}_{\text{tr}}) = \int [-j\omega\mu_0 \cdot \boldsymbol{J}_{\text{tar}}(\boldsymbol{r}_{\text{tar}}) \cdot G(\boldsymbol{r}_{\text{tr}}, \boldsymbol{r}_{\text{tar}}) \\ + \boldsymbol{M}_{\text{tar}}(\boldsymbol{r}_{\text{tar}}) \times G(\boldsymbol{r}_{\text{tr}}, \boldsymbol{r}_{\text{tar}})] e^{j\varphi_s} \cdot T_{\parallel 0 \to 1}^{S} \\ \cdot T_{\parallel 1 \to 0}^{S} \, \mathrm{d} \boldsymbol{r}_{\text{tar}} \quad (18)$$

where the $\boldsymbol{r}_{\text{tar}}$ is arbitrary point on the target. $\boldsymbol{J}_{\text{tar}}(\boldsymbol{r}_{\text{tar}})$ and $\boldsymbol{M}_{\text{tar}}(\boldsymbol{r}_{\text{tar}})$ are the equivalent electric current and equivalent magnetic current at $\boldsymbol{r}_{\text{tar}}$ on the target, respectively. They are calculated as:

$$\boldsymbol{J}_{\text{tar}}(\boldsymbol{r}_{\text{tar}}) = (1 - R_{\text{tar}}) \boldsymbol{n} \times \boldsymbol{H}_i(\boldsymbol{r}_{\text{tar}}) \quad (19)$$

$$\boldsymbol{M}_{\text{tar}}(\boldsymbol{r}_{\text{tar}}) = (1 + R_{\text{tar}}) \boldsymbol{E}_i(\boldsymbol{r}_{\text{tar}}) \times \boldsymbol{n} \quad (20)$$

where $\boldsymbol{E}_i(\boldsymbol{r}_{\text{tar}})$ and $\boldsymbol{H}_i(\boldsymbol{r}_{\text{tar}})$ are the incident fields propagating at $\boldsymbol{r}_{\text{tar}}$ of the target. $\boldsymbol{E}_i(\boldsymbol{r}_{\text{tar}})$ is approximated as follows:

$$\boldsymbol{E}_i(\boldsymbol{r}_{\text{tar}}) = \boldsymbol{E}_0 H_0^{(2)}(k|\boldsymbol{r}_{\text{tr}} - \boldsymbol{r}_{\text{tar}}|) e^{j\varphi_s} \cdot T_{\parallel 0 \to 1}^{i} \cdot T_{\parallel 1 \to 0}^{i} \quad (21)$$

$R_{\text{tar}}$ is the reflection coefficient of the target which is calculated as:

$$R_{\text{tar}} = \frac{\cos \theta_1^i(\boldsymbol{r}_{\text{tar}}) - \sqrt{\frac{\varepsilon_2}{\varepsilon_0}} \sqrt{1 - \left(\frac{\varepsilon_0}{\varepsilon_2}\right) \sin^2 \theta_1^i(\boldsymbol{r}_{\text{tar}})}}{\cos \theta_1^i(\boldsymbol{r}_{\text{tar}}) + \sqrt{\frac{\varepsilon_2}{\varepsilon_0}} \sqrt{1 - \left(\frac{\varepsilon_0}{\varepsilon_2}\right) \sin^2 \theta_1^i(\boldsymbol{r}_{\text{tar}})}} \quad (22)$$

where $\varepsilon_2$ is the complex permittivity of the target. $\theta_1^i(\boldsymbol{r}_{\text{tar}})$ is the incident angle of the wave at $\boldsymbol{r}_{\text{tar}}$ of the target.

The total received signal by the transceiver at the position $\boldsymbol{r}_{tr}$ is estimated by adding the reflections from the obstacle and target:

$$S(\boldsymbol{r}_{\text{tr}}) = \boldsymbol{E}_{\text{obs}}(\boldsymbol{r}_{\text{tr}}) + \boldsymbol{E}_{\text{tar}}(\boldsymbol{r}_{\text{tr}}) \quad (23)$$

Since the wideband FMCW radar is utilized, the simulation is conducted at different frequencies to model the FMCW radar. Finally, the simulated signal dataset is denoted as:

$$\boldsymbol{S} = \begin{cases} [S(\boldsymbol{r}_1, \omega_1), \dots, S(\boldsymbol{r}_p, \omega_1), \dots, S(\boldsymbol{r}_P, \omega_1)] \\ \vdots \\ [S(\boldsymbol{r}_1, \omega_q), \dots, S(\boldsymbol{r}_p, \omega_q), \dots, S(\boldsymbol{r}_P, \omega_q)] \\ \vdots \\ [S(\boldsymbol{r}_1, \omega_Q), \dots, S(\boldsymbol{r}_p, \omega_Q), \dots, S(\boldsymbol{r}_P, \omega_Q)] \end{cases} \quad (24)$$

where $\boldsymbol{r}_p$ is the *p*-th measured position and $\omega_q$ is the *q*-th sampled frequency in the FMCW radar. *P* and *Q* are the total measurement positions and sampled frequencies, respectively.

### III. MEASUREMENT SYSTEM WITH FMCW RADAR AND SIMULATION CONFIGURATION

#### A. Measurement System

In this work, the mmWave FMCW radar is utilized for non-destructive subsurface imaging. mmWave FMCW radar has been widely utilized in many sensing scenarios [23]. Compared with lower-frequency signals such as Wi-Fi [24, 25], the FMCW radar has been miniaturized and mass-produced, and it also has the advantages of portability and low cost. Since mmWave FMCW radar emits wideband electromagnetic waves, it is able to separate objects at different distances. Thus, it is possible to detect the positions of obstacles such as the walls of building and the target shadowed by the obstacles simultaneously. The radar used to build the measurement system operates at 79 GHz band and the bandwidth is 3.6 GHz from 77.2 ~ 80.8 GHz (T14, S-Takaya). It is a COTS module and based on TI-IWR1443 chipset.

The measurement system is developed as shown in Fig. 3. The radar is mounted on the base of a slider and the slider is utilized to move the radar. A two-phase stepping motor is connected to the slider to precisely control the movement increment. A motor driver is utilized to drive the motor by sending powerful pulses. The driver is controlled by a one-board microcomputer (Arduino, UNO). The control PC is connected to both the microcomputer and the radar. The PC sends commands to both microcomputer and radar to coordinate the operation of the whole system. During experiment, the measurement is conducted at 1 mm step. Each time the radar completes the data acquisition and signaling will be sent back to the PC and PC will then send the command to trigger the movement of the slider with 1 mm. At the next position, the radar measurement will be repeated. With this procedure, the scanning is conducted within 250 mm range on the *x* axis.

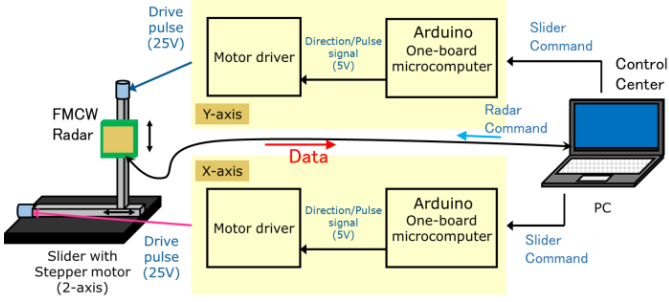

Fig. 3. Schematic diagram of GO for simulating wave transmission inside the obstacle.

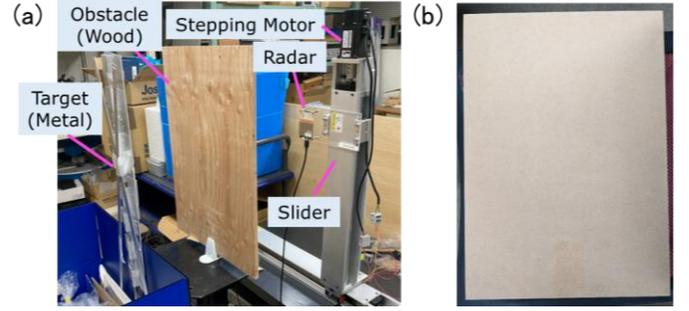

Fig. 4. (a) Measurement system with softwood obstacle and the metallic target. (b) Medium density fiberboard.

All the data is saved in the PC for offline processing. In general, the instantaneous frequency of the radiated FMCW signal is expressed by the following:

$$f(t) = f_0 + Kt \quad (25)$$

where $f_0$ is the carrier frequency, $K$ is the frequency slope, and $t$ is time. FMCW signal can be represented as:

$$s_T(t) = A_0 \cos\left(2\pi\left(f_0 t + \frac{K}{2}t^2\right) + \varphi_0\right) \quad (26)$$

where $A_0$ and $\varphi_0$ are the amplitude and initial phase of the transmitted signal, respectively. When the FMCW signal encounters the objects, it will be reflected and received by the radar. Denote the reflected signal as:

$$s_R(t) = A_0 \sigma \cos\left(2\pi\left(f_0(t-\tau) + \frac{K}{2}(t-\tau)^2\right) + \varphi_0\right) \quad (27)$$

where $\sigma$ is the attenuation factor, and $\tau$ is signal delay between the transmitted and reflected signals. Then, the reflected signal is mixed with the transmitted signal, and the intermediate frequency (IF) signal is obtained by:

$$s_{\mathrm{IF}} = \mathrm{LPF}(s_T \cdot s_R) = \frac{A_0^2}{2}\sigma \cos(2\pi\tau(f_0 + Kt)) \quad (28)$$

Here LPF means low pass filter. Since the utilized radar module is based on complex-baseband architecture, the IF signal is sampled in both in-phase and quadrature paths. Then, the digitized and saved signal can be expressed as:

$$s_{\mathrm{IF}}[n] = \frac{A_0^2}{2}\sigma e^{j2\pi\tau(Kn\Delta t + f_0)} \quad (29)$$

where $\Delta t$ is the sampling interval in the radar. $s_{\mathrm{IF}}[n]$ is utilized for imaging.

During the measurement, the wood is utilized as the obstacle and the metal was utilized as the target as shown in Fig. 4 (a). The distance between the radar and wood obstacle is around 250 mm and the distance between the radar and the target is around 500 mm. In the experiment, two different wood boards were used. One is softwood plywood which is the obstacle shown in Fig. 4(a). The other is the medium density fiberboard (MDF) as shown in Fig. 4(b). It can be observed that the patterns of the two boards are different. There are wood grains on the softwood board. While on the MDF board, there is no significant pattern since it is made from wood fibers after pulping. The purpose of using different woods is to investigate the influence of the obstacle pattern on the imaging results. It is also utilized to verify the performance of the proposed simulation method in modeling the FMCW radar subsurface imaging under different obstacles.

### B. Simulation Configuration

The simulation is configured to mimic the subsurface imaging of metallic target behind the wood obstacle as depicted in the previous Section III.A. The simulation structure is the same as shown in Fig. 1. The complex permittivity of obstacle $\varepsilon_1$ and target $\varepsilon_2$ are defined as 1.99 - j1.12 and 1 - j2.28×10$^6$, respectively, which is obtained from the MATLAB building material database. Since the utilized radar operates at 77.2 ~ 80.8 GHz and the sampling point is 256, the simulation is carried in the same bandwidth with the frequency step of 14.0625 MHz. Regarding the measurement setup, after the simulation is completed at one measurement point, the transmitter is moved with 1 mm increment on the $x$ axis. The simulation is also repeated within 250 mm range. After all the simulation data are obtained, the synthesis aperture radar (SAR) processing is applied to generate the image [26].

## IV. SIMULATION AND MEASUREMENT RESULTS

### A. Imaging without Obstacles

In the evaluation, the target with the length of 100 mm is utilized. First, the scenario where there is no obstacle and only the metallic target exists is considered. The simulation and measurement results are shown in Fig. 5. The color bar is normalized to the maximum in the reconstructed image. It can be observed that the metallic object is successfully reconstructed in the image. Choose the center of the target image on the $y$ axis, the intensity distribution on the $x$ axis can be depicted as Fig. 6. The full width at half maximum (FWHM) is utilized to estimate the length of image. The estimated lengths of the target in the simulation and measurement results are 99.7 mm and 94.7 mm, respectively, demonstrating the correctness of the imaging results. The slight difference in the alignment on $x$ axis is caused by the physical displacement of the object in measurement. It can be noted that the target image has thickness on the $y$ axis although the metal reflects all the incident waves at the surface. This is because the bandwidth of the radar is only 3.6 GHz, the range resolution is only 41.7 mm. Therefore, the image will be shown as a tube with thickness of around 40 mm even though the object is metal. With the current system configuration, targets whose distance is less than 41.7 mm are difficult to differentiate. Compared with simulation, the position on *y* axis in the measurement results is larger than the set position. This is considered to be caused by the delay inside the radar module. These results demonstrate that the proposed simulation method can model the wave propagation and generate the dataset for SAR imaging.

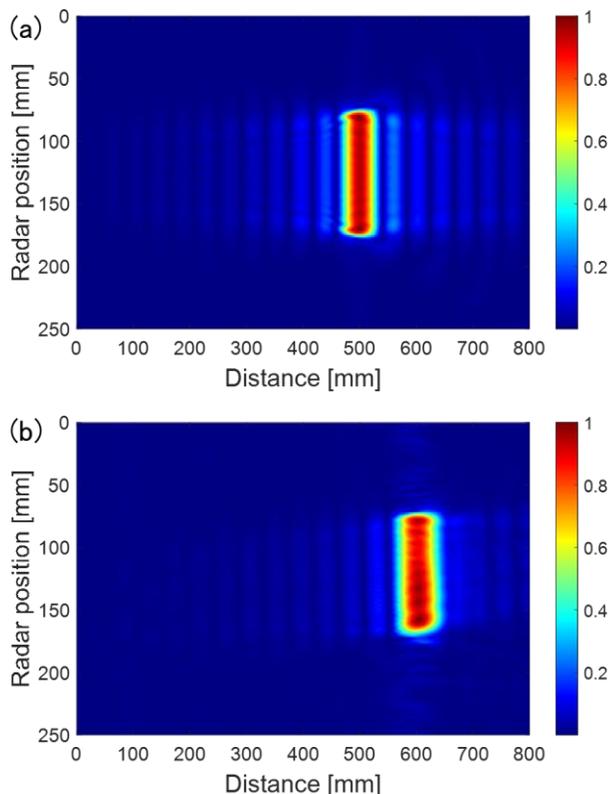

Fig. 5. Imaging result of a metallic target without any obstacles. (a) Simulation and (b) Measurement.

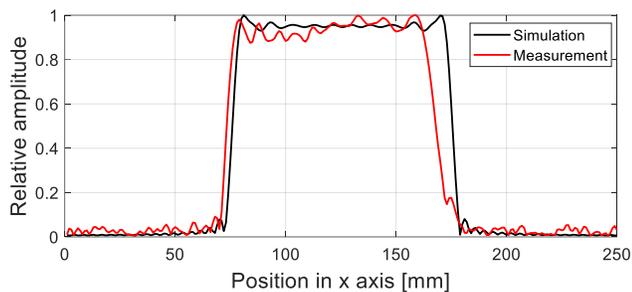

Fig. 6. Intensity distribution along the *x* axis at the center of target image when there is no obstacle.

### B. Subsurface Imaging with MDF as Obstacle

The MDF wood was utilized as the obstacle which was placed between the metallic target and the radar. The thickness of the MDF is 9 mm. In the simulation, the thickness of the obstacle *l* was set the same as the real one. And the MDF was modeled as a homogenous object with the complex permittivity of $\varepsilon_1$ as defined in Section III.B. In the measurement, the experiment procedure was the same as the one in Section IV.A. The imaging results by utilizing the dataset from simulation and measurement are shown in Fig. 7. The color bar is normalized to the maximum in the reconstructed image. From the simulation result, it can be observed that both the obstacle and target are successfully reconstructed in the image. On the other hand, both objects are also recognized in the measurement results. The intensity distribution on the *x* axis at the center of target image is shown in Fig. 8. The intensity was normalized to the maximum intensity in the case when there was no obstacle as shown in Fig. 6. The estimated lengths of the target in the simulation and measurement results are 99.1 mm and 94.2 mm, respectively, which reflects the size of the target. Further, by comparing the intensity of the target image, it can be found that the intensity in measurement shows good agreement with that in the simulation. This indicates that the attenuation effect of the obstacle is successfully modeled with the proposed simulation method. These results also demonstrate that the subsurface imaging is efficient under the condition that the MDF board is used. By comparing the image intensity of the obstacle, it can be observed that the intensity in measurement is weaker than that in the simulation. This may be caused by the roughness of the real MDF board. Since the imaging algorithm assumes that the surface is smooth, the roughness in the measurement degrades the image quality. This may also be caused by the mismatch of the permittivity in the simulation and measurement, which makes the reflection from the obstacle different.

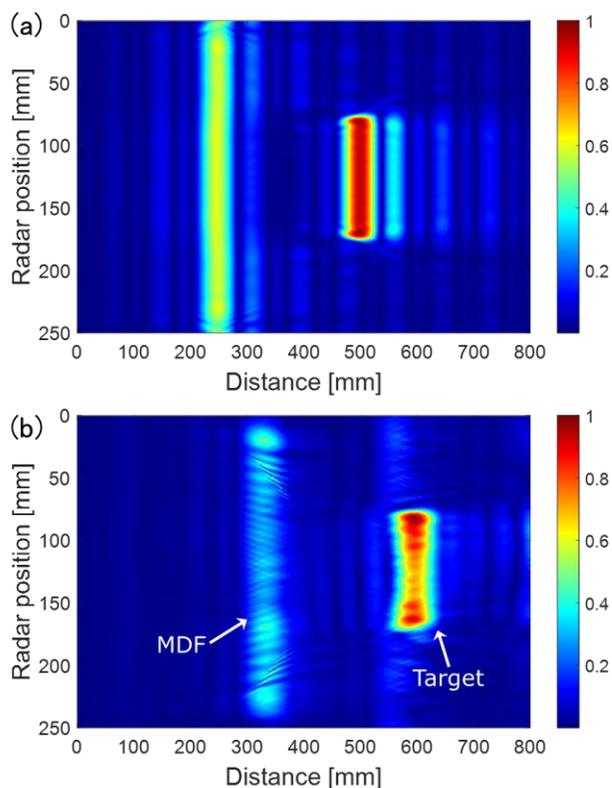

Fig. 7. Imaging result of a metallic target with MDF obstacle. (a) Simulation and (b) Measurement.

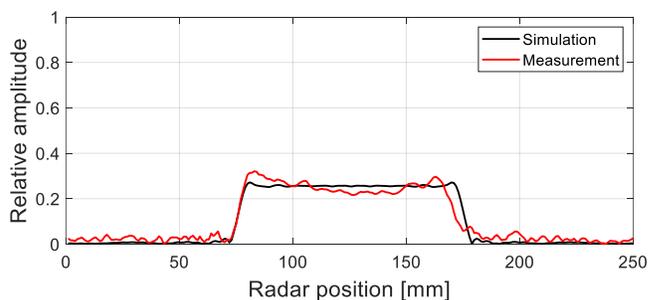

Fig. 8. Intensity distribution along the *x* axis at the center of target image when the obstacle is MDF. The intensity is normalized to the case when there is no obstacle.

*C. Subsurface Imaging with Softwood as Obstacle*

The softwood plywood was utilized as the obstacle to be placed between the radar and target. The thickness of the softwood is 10 mm. The softwood has patten as shown in Fig. 4(a) and these patterns should have different characteristics. Therefore, in the simulation, the obstacle was modeled as inhomogeneous. When the density of the wood increases, the permittivity is also supposed to increase. Thus, except for the wood permittivity $\varepsilon_1$, another permittivity $\varepsilon_3 = 2.5 - j0.2$ was utilized to depict the inhomogeneity of the obstacle. The model of softwood in simulation is configured to have a periodic pattern, whose permittivity was set to change every 10-mm length between $\varepsilon_1$ and $\varepsilon_3$.

The target was not changed. The measurement was conducted with the same procedure as that of Section IV.B. The imaging results from the simulation and measurement are shown in Fig. 9. The color bar is normalized to the maximum in the reconstructed image. From the measurement result, it can be observed that the images of the obstacles and target are not continuous. This is considered to be caused by the inhomogeneity of the softwood and the patterns were reflected in the image. From the simulation results, it can be seen that similar images are reconstructed. The image of the target is also not continuous but the intensity changes dynamically. Figure 10 shows the intensity distribution on the *x* axis at the center of target image. The intensity was normalized to the maximum intensity of no obstacle case, but the scale is set differently to better show the fluctuation of the intensity. The estimated lengths of the target in the simulation and measurement results are 97.3 mm and 91.9 mm, respectively. It can be observed that the image quality degrades under the blocking of softwood. Due to the inhomogeneity of obstacle, the image of the target is also influenced. From the simulation result, the image intensity fluctuates where the distance between peaks is around 20 mm. Similarly, in the measurement result, the distance between peaks in the intensity is also around 20 mm. And the overall intensity shows good agreement between simulation and measurment, demonstrating that the simulation modeled the measurement well in the obstacle attenuation. These results indicate that the proposed simulation method is able to model the inhomogeneity of the obstacle in the subsurface imaging.

*D. Discussion*

From the measurement, it can be observed that the detection of an object becomes difficult when there is a dielectric obstacle between the radar and the object. The difficulty is due to not only the attenuation of the obstacle, but also the inhomogeneity of the obstacle. In the MDF case, the intensity of the target image is only around 30% of that in the no-obstacle case. In the softwood case, the intensity of the targe image further decreases, and the image shows large fluctuation although the target is a smooth metal plate. When the obstacle is inhomogeneous, the wave propagates through different points and encounters various attenuation conditions. The parts with high dielectric constant and conductivity can weaken the incident wave from the radar and the wave arrives at the target with lower intensity. These parts will also weaken the scattered wave from the shielded object. Therefore, in the image, some parts are with low intensity while some are relatively high.

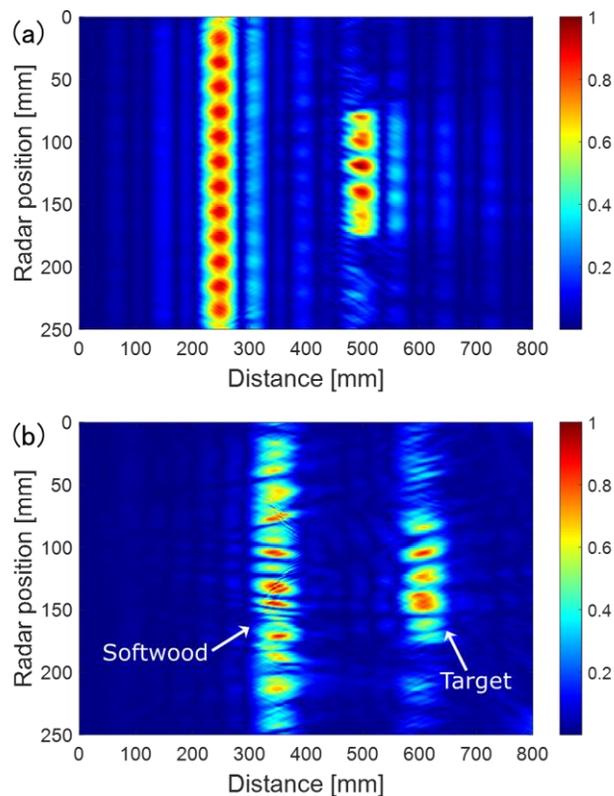

Fig. 9. Imaging result of a metallic target with softwood obstacle. (a) Simulation and (b) Measurement.

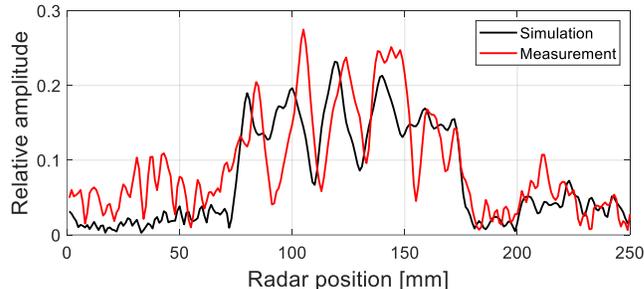

Fig. 10. Intensity distribution along the *x* axis at the center of target image when the obstacle is softwood. The intensity is normalized to the case when there is no obstacle.

In the simulation, for MDF obstacle case, the intensity of the target image is also around 30% of that in the no-obstacle case, demonstrating that the permittivity utilized can mimic the attenuation effect of the MDF use in the experiment. In the softwood obstacle simulation, similar image fluctuation is produced by adding the inhomogeneity to the obstacle. This agreement with the measurement indicates that the analysis of the cause of image fluctuation in softwood measurement is meaningful. From all the results in different scenarios corresponding to the measurement conditions, it can be found that the proposed method can model the reflections from both target and obstacle, the attenuation effect of obstacle, and the influence of inhomogeneous obstacle for subsurface imaging with mmWave FMCW radar.

## V. CONCLUSION

The HPGO method was proposed to model the subsurface imaging using mmWave FMCW radar. In HPGO, the physical optics method was utilized to calculate the reflection from obstacle and target. And geometrical optics method was utilized to calculate the transmission of the wave through the obstacle. The experiment setup was developed to evaluate the performance of the HPGO method. The simulations and corresponding measurements were conducted under different conditions. The image results showed good agreement between the simulation and measurement, demonstrating the efficacy of HPGO method in modeling the reflections, attenuation, and inhomogeneity factors for the subsurface imaging. By using the proposed method, the imaging results can be better interpreted.

## REFERENCES

[1] J. -T. Gonzalez-Partida, P. Almorox-Gonzalez, M. Burgos-Garcia, B. -P. Dorta-Naranjo and J. I. Alonso, "Through-the-Wall Surveillance With Millimeter-Wave LFMCW Radars," *IEEE Transactions on Geoscience and Remote Sensing*, vol. 47, no. 6, pp. 1796-1805, 2009.
[2] A. A. Pramudita *et al.*, "Radar System for Detecting Respiration Vital Sign of Live Victim Behind the Wall," in *IEEE Sensors Journal*, vol. 22, no. 15, pp. 14670-14685, 2022.
[3] S. Kharkovsky and R. Zoughi, "Microwave and millimeter wave nondestructive testing and evaluation - Overview and recent advances," *IEEE Instrumentation and Measurement Magazine*, vol. 10, no. 2, pp. 26-38, 2007.
[4] B. Gonzalez-Valdes, Y. Alvarez, S. Mantzavinos, C. M. Rappaport, F. Las-Heras and J. A. Martinez-Lorenzo, "Improving Security Screening: A Comparison of Multistatic Radar Configurations for Human Body Imaging," *IEEE Antennas and Propagation Magazine*, vol. 58, no. 4, pp. 35-47, 2016.
[5] X. Tao, D. Zhang, Z. Wang, X. Liu, H. Zhang, and D. Xu, "Detection of Power Line Insulator Defects Using Aerial Images Analyzed With Convolutional Neural Networks," *IEEE Transactions on Systems, Man, and Cybernetics: Systems*, vol. 50, no. 4, pp. 1486-1498, 2020.
[6] A. Hirata, K. Suizu, Y. Sudo, I. Watanabe, N. Sekine and A. Kasamatsu, "Non-destructive Inspection of Concrete Surface Crack Using Near-Field Scattering," *2020 IEEE International Symposium on Radio-Frequency Integration Technology (RFIT)*, Hiroshima, Japan, 2020, pp. 244-246.
[7] A. Aljurbua and K. Sarabandi, "Detection and Localization of Pipeline Leaks Using 3-D Bistatic Subsurface Imaging Radars," in *IEEE Transactions on Geoscience and Remote Sensing*, vol. 60, pp. 1-11, 2022, Art no. 5220211.
[8] M. T. Ghasr, S. Kharkovsky, R. Bohnert, B. Hirst and R. Zoughi, "30 GHz Linear High-Resolution and Rapid Millimeter Wave Imaging System for NDE," *IEEE Transactions on Antennas and Propagation*, vol. 61, no. 9, pp. 4733-4740, 2013
[9] Y. Gao and R. Zoughi, "Millimeter Wave Reflectometry and Imaging for Noninvasive Diagnosis of Skin Burn Injuries," *IEEE Transactions on Instrumentation and Measurement*, vol. 66, no. 1, pp. 77-84, 2017
[10] M. Becquaert, E. Cristofani, B. Lauwens, M. Vandewal, J. H. Stiens and N. Deligiannis, "Online Sequential Compressed Sensing With Multiple Information for Through-the-Wall Radar Imaging," *IEEE Sensors Journal*, vol. 19, no. 11, pp. 4138-4148, 2019.
[11] B. Yektakhah and K. Sarabandi, "All-Directions Through-the-Wall Imaging Using a Small Number of Moving Omnidirectional Bi-Static FMCW Transceivers," *IEEE Transactions on Geoscience and Remote Sensing*, vol. 57, no. 5, pp. 2618-2627, 2019.
[12] A. A. Pramudita *et al.*, "Radar System for Detecting Respiration Vital Sign of Live Victim Behind the Wall," *IEEE Sensors Journal*, vol. 22, no. 15, pp. 14670-14685, 2022.
[13] S. M. H. Naghavi, M. T. Taba, M. Aseeri and E. Afshari, "An Integrated 100-GHz FMCW Imaging Radar for Low-Cost Drywall Inspection," *IEEE Transactions on Microwave Theory and Techniques*, vol. 72, no. 2, pp. 1070-1084, 2024
[14] Z. Tian, J. Zhang and S. Yi, "Innovative W-Band Through-Wall Radar With Sector Scanning: Utilizing Traveling Wave Tubes for Enhanced Penetration," *IEEE Sensors Journal*, vol. 24, no. 19, pp. 30801-30809, 2024.
[15] M. E. Yanik, D. Wang and M. Torlak, "Development and Demonstration of MIMO-SAR mmWave Imaging Testbeds," in *IEEE Access*, vol. 8, pp. 126019-126038, 2020.
[16] A. Hirata, M. Nakashizuka, K. Suizu and Y. Sudo, "Improvement of Detection in Concrete Surface Cracks Covered with Paper by Using Standing Wave of 77-GHz-Band Millimeter-Wave," *2019 IEEE MTT-S International Microwave Symposium (IMS)*, Boston, MA, USA, 2019, pp. 297-300.
[17] L. Liu, Z. Liu and B. E. Barrowes, "Through-Wall Bio-Radiolocation With UWB Impulse Radar: Observation, Simulation and Signal Extraction," *IEEE Journal of Selected Topics in Applied Earth Observations and Remote Sensing*, vol. 4, no. 4, pp. 791-798, 2011.
[18] J. Helander, A. Ericsson, M. Gustafsson, T. Martin, D. Sjöberg and C. Larsson, "Compressive Sensing Techniques for mm-Wave Nondestructive Testing of Composite Panels," *IEEE Transactions on Antennas and Propagation*, vol. 65, no. 10, pp. 5523-5531, 2017
[19] X. Zhang, J. Liang, N. Wang, T. Chang, Q. Guo and H. -L. Cui, "Broadband Millimeter-Wave Imaging Radar-Based 3-D Holographic Reconstruction for Nondestructive Testing," in *IEEE Transactions on Microwave Theory and Techniques*, vol. 68, no. 3, pp. 1074-1085, March 2020.
[20] A. Aljurbua and K. Sarabandi, "Detection and Localization of Pipeline Leaks Using 3-D Bistatic Subsurface Imaging Radars," *IEEE Transactions on Geoscience and Remote Sensing*, vol. 60, pp. 1-11, 2022.
[21] S. Eide, T. Casademont, Ø. L. Aardal and S. -E. Hamran, "Modeling FMCW Radar for Subsurface Analysis," *IEEE Journal of Selected Topics in Applied Earth Observations and Remote Sensing*, vol. 15, pp. 2998-3007, 2022.
[22] C. A. Balanis, "Advanced Engineering Electromagnetics," Wiley, 2012.
[23] A. Venon, Y. Dupuis, P. Vasseur and P. Merriaux, "Millimeter Wave FMCW RADARs for Perception, Recognition and Localization in Automotive Applications: A Survey," *IEEE Transactions on Intelligent Vehicles*, vol. 7, no. 3, pp. 533-555, 2022.
[24] H. Song, B. Wei, Q. Yu, X. Xiao and T. Kikkawa, "WiEps: Measurement of Dielectric Property With Commodity WiFi Device—An Application to Ethanol/Water Mixture," *IEEE Internet of Things Journal*, vol. 7, no. 12, pp. 11667-11677, 2020.
[25] B. Wei, H. Song, J. Katto and T. Kikkawa, "RSSI–CSI Measurement and Variation Mitigation With Commodity Wi-Fi Device," *IEEE Internet of Things Journal*, vol. 10, no. 7, pp. 6249-6258, 2023.
[26] J. M. Lopez-Sanchez and J. Fortuny-Guasch, "3-D radar imaging using range migration techniques," *IEEE Transactions on Antennas and Propagation*, vol. 48, no. 5, pp. 728-737, 2000.